\documentclass[%
 reprint,
 superscriptaddress,
preprintnumbers,
nofootinbib,
 amsmath,amssymb,
prd,
floatfix,
longbibliography
]{revtex4-2}\usepackage[utf8]{inputenc}
\usepackage{bm}
\usepackage{mathbbol}
\usepackage{graphicx}
\usepackage{hyperref}
\usepackage{orcidlink}

\newcommand{\eg}{\textit{e.g.}}
\newcommand{\ie}{\textit{i.e.}}

\begin{document}
\preprint{LA-UR-25-28522}

\title{Once-in-a-lifetime encounter models for neutrino media II:\\
Quasi-steady states and miscidynamic flavor evolution}

\newcommand{\LANL}{Theoretical Division, Los Alamos National Laboratory, Los Alamos, New Mexico 87545, USA}
\newcommand{\UNM}{Department of Physics and Astronomy, University of New Mexico, Albuquerque, New Mexico 87131, USA}

\author{Anson Kost\,\orcidlink{0009-0008-0327-5057}}

\affiliation{\UNM}

\author{Lucas Johns}
\affiliation{\LANL}

\author{Huaiyu Duan\,\orcidlink{0000-0001-6708-3048}}
\affiliation{\UNM}
\date{\today}

\begin{abstract}
We extended the once-in-a-lifetime encounter (OILE) model to stochastic interactions among neutrinos. As in the original OILE model, the new model reproduces the mean-field behavior of a dense neutrino gas for time $t\lesssim (\mu\gamma)^{-1}$, where $\mu$ measures the strength of the mean-field neutrino self-interaction potential and is proportional to the neutrino density, and the dimensionless ``impact parameter'' $\gamma$ is a measure of the change in the flavor quantum state of a neutrino during interaction with another neutrino when the wave packets of the two neutrinos overlap. As in the mean-field case, the OILE model with random neutrino velocities experiences kinetic flavor decoherence as the flavor quantum states of the neutrinos diverge from each other. Unlike the mean-field case, however, the OILE model has a ``collision term'' due to the quantum entanglement among neutrinos. For $\gamma\ll1$, this incoherent effect can drive the neutrinos into a quasi-steady state that is similar to the collective precession mode in a homogeneous and isotropic neutrino gas in the mean-field approximation. Subsequently, the collision term drives the neutrino gas adiabatically through different quasi-steady states and eventually to flavor equilibration. This process is an example of miscidynamic flavor evolution, with the mixing equilibria being the quasi-steady precession states.
\end{abstract}

\maketitle

\section{Introduction}
The flavor quantum state of a neutrino can be altered through refraction as it propagates through a dense medium. A famous example is the Mikheyev-Smirnov-Wolfenstein (MSW) mechanism in which the electron-flavor and muon / tau-flavor neutrinos have different amplitudes of forward scattering by ambient electrons and thus different refractive indices in matter \cite{Wolfenstein:1977ue,Mikheyev:1985zog}. Similarly, neutrino-neutrino forward scattering in a dense neutrino medium in, \eg, the early Universe, a core-collapse supernova (CCSN) or a neutron-star merger (NSM), can also give rise to different refractive indices for neutrinos of different flavors \cite{Fuller:1987aa,Notzold:1987ik}. Unlike the MSW mechanism, however, neutrino-neutrino forward-scattering, aka neutrino self-interactions, also allows neutrinos to exchange flavors with each other \cite{Pantaleone:1992eq}, leading to many interesting nonlinear flavor evolution phenomena that are still under intense investigation (see \cite{Duan:2010bg,Chakraborty:2016yeg,Tamborra:2020cul,Volpe:2023met,Johns:2025mlm} for some reviews).

In the weak inhomogeneity limit where the length scales of the inhomogeneities are much larger than the de Broglie wavelengths of the neutrinos, the flavor content of a neutrino gas can be described by the Wigner distribution or the flavor (density) matrix with elements \cite{Sigl:1993ctk}:
\begin{align}
    [\mathsf{F}_{\vec{p}}]_{\alpha\beta} = 
    \int\frac{\mathrm{d}^3 q}{(2\pi)^3} \, e^{\mathrm{i} \vec{q} \cdot \vec{r}} \,
    \left\langle a_\beta^\dagger\left(t, \vec{p} -\frac{\vec{q}}{2}\right) a_\alpha\left(t, \vec{p} + \frac{\vec{q}}{2}\right) \right\rangle,
    \label{eq:flavor-matrix}
\end{align}
where $a_\alpha^\dagger (t, \vec{k})$ and $a_\alpha(t, \vec{k})$ are the creation and annihilation operators, respectively, of the neutrino of flavor $\alpha$ with momentum $\vec{k}$ in the Heisenberg picture. The diagonal elements of $\mathsf{F}_{\vec{p}}(t, \vec{r})$ give the number densities of the neutrinos of the corresponding flavor at time $t$ and position $\vec{r}$, and its off-diagonal elements encode the flavor coherence. The flavor evolution of the neutrino gas is governed by the quantum kinetic equation (QKE) \cite{Sigl:1993ctk}:
\begin{align}
    (\partial_t + \vec{v} \cdot \vec{\nabla}) \mathsf{F}_{\vec{p}} = -\mathrm{i} [\mathsf{H}_{\vec{p}}, \mathsf{F}_{\vec{p}}] + \mathsf{C}_{\vec{p}},
    \label{eq:qke}
\end{align}
where $\vec{v} = \vec{p}/|\vec{p}|$ is the velocity of the neutrino, and the commutator and $\mathsf{C}_{\vec{p}}$ account for coherent processes and incoherent ``collisions'' (including emissions and absorptions) of the neutrinos, respectively. For simplicity, we adopt the two-flavor mixing scheme in this paper with the nominal neutrino flavors $e$ and $\tau$. In this scheme, the flavor matrix can be expressed in terms of the (flavor) polarization vector through the Pauli matrices $\sigma_i$ ($i=x,y,z$),
\begin{align}
    \bm{F}_{\vec{p}} = \mathrm{tr}( \bm{\sigma}\mathsf{F}_{\vec{p}}),
\end{align}
which obeys the equation of motion
\begin{align}
    \partial_t \bm{F}_{\vec{p}} + \vec{v} \cdot \vec{\nabla} \bm{F}_{\vec{p}} = \bm{H}_{\vec{p}} \times \bm{F}_{\vec{p}} + \bm{C}_{\vec{p}}.
    \label{eq:qke-P}
\end{align}
Throughout this paper, we use $x$, $y$, and $z$ to denote the three orthogonal directions in flavor space. In the flavor basis, a polarization vector in the $+z$ ($-z$) direction represents a neutrino in the $e$ ($\tau$) flavor. The QKE without the collision term is also known as the mean-field approximation in which the neutrinos are described by a classical field without quantum entanglement. The effects of weak quantum correlations among neutrinos can be captured by a collision term in the QKE \cite{Blaschke:2016xxt,Froustey:2020mcq}.

Whether the single-particle picture of Eq.~\eqref{eq:qke} is sufficient to describe the nonlinear flavor evolution of a dense neutrino medium was questioned very early on \cite{Pantaleone:1992eq}. It has since been debated and investigated by independent groups, and it has drawn more attention recently because of the rising interest in quantum computing (\eg, \cite{Bell:2003mg,Friedland:2003eh,Cervia:2019res,Patwardhan:2021rej,Roggero:2021asb,Martin:2021bri,Illa:2022zgu,Martin:2023ljq,Martin:2023gbo}; see also \cite{Patwardhan:2022mxg} for a review devoted to this subject). In contrast to the QKE, a many-body model commonly used in the literature for neutrino oscillations is a closed system of $N$ neutrinos with definite momenta that interact through the forward scattering Hamiltonian \cite{Bell:2003mg,Friedland:2003eh}
\begin{align}
    H_{\nu\nu} = \sum_{a < b}^N V_{ab}
    \label{eq:Hvv}
\end{align}
with
\begin{align}
    V_{ab} = \frac{G_\mathrm{F}}{\sqrt2 \mathcal{V}} (1 - \vec{v}_a\cdot\vec{v}_b) \sum_{j=x,y,z}\sigma_j^{(a)} \otimes \sigma_j^{(b)},
    \label{eq:Vab}
\end{align}
where $G_\mathrm{F}$ is the Fermi coupling constant, $\mathcal{V}$ is the quantization volume, and $a$ and $b$ are the indices of the interacting neutrinos. It has recently been shown that $H_{\nu\nu}$ without any artificially imposed symmetry is nonintegrable and can lead to flavor equilibration through quantum entanglement among neutrinos \cite{Martin:2023ljq,Martin:2023gbo}.

The applicability of this many-body model to astrophysical environments such as CCSNe and NSMs has also been questioned in several ways (\eg, \cite{Johns:2023ewj,Shalgar:2023ooi}). One criticism is that Eq.~\eqref{eq:Hvv} leaves out the momentum-changing part of the $\nu$-$\nu$ interaction. However, a preliminary study suggests that inclusion of the momentum-changing interaction only accelerates the flavor equilibration \cite{Cirigliano:2024pnm}. Another criticism is that the neutrinos in the many-body model interact indefinitely, while neutrinos in astrophysical environments can interact only briefly. This is because neutrinos in these environments have de Broglie wavelengths ($\sim 10$~fm) that are much larger than that of the $Z$ boson at rest ($\sim 10^{-3}$~fm), and, as a result, neutrinos can interact only when their wave packets overlap with each other. It is not obvious whether the wave packet picture can agree with either the QKE or the aforementioned many-body model. For the former, this picture breaks the weak inhomogeneity assumption, and the flavor matrix in Eq.~\eqref{eq:flavor-matrix} and the QKE need to be coarse-grained over scales larger than the size of the neutrino wave packet $\sigma_\text{wp}$. For the latter, many momentum modes are needed to construct a wave packet, while the computational cost of the many-body model typically scales exponentially with the number of modes.

In an earlier paper \cite{Kost:2024esc}, we proposed the once-in-a-lifetime encounter (OILE) model in which any pair of neutrinos can interact only briefly and at most once in their lifetimes. Under this assumption, it is possible to describe the flavor evolution of each neutrino with its one-particle (flavor) density operator $\rho$ or, equivalently, its (flavor) Bloch vector $\bm{P} = \mathrm{tr}(\bm{\sigma} \rho )$, except while it is interacting with another neutrino. Using the OILE model, we demonstrated that mean-field behavior can emerge from the particle picture in one limit and flavor equilibration can be achieved in the opposite limit. For example, it can be shown that the flavor evolution in the model of a neutrino interacting with a uniform medium of neutrinos of the same flavor is governed by the following master equation (see Sec.~\ref{sec:OILE2} for details):
\begin{align}
    \frac{\mathrm{d}\bm{P}}{\mathrm{d}t} = (-\omega\bm{B} + \mu\bm{Q})\times\bm{P} + \mu\gamma(\bm{Q} - \bm{P}),
    \label{eq:uniform-bg}
\end{align}
where $\omega$ is the vacuum oscillation frequency of the foreground neutrino, $\bm{B} = (0,0,1)$ in the mass basis, $\bm{Q}$ is the Bloch vector of the background neutrinos, $\gamma=\sqrt{2}G_\mathrm{F}\delta t/\mathcal{V}$ is the dimensionless ``impact parameter'' that measures the change of the flavor quantum state of the foreground neutrino during the interaction interval $\delta t$, and $\mu = \gamma \mathcal{P} / \delta t = \sqrt{2}G_\mathrm{F} n_\nu$ is proportional to the probability of interaction $\mathcal{P}$ during the interval $\delta t$ (with $\mathcal{V}$ held fixed), or, equivalently, the density of the background neutrinos $n_\nu$. Equation~\eqref{eq:uniform-bg} obviously takes the form of Eq.~\eqref{eq:qke-P}. It reproduces mean-field behavior at $t\lesssim (\gamma \mu)^{-1}$. However, at $t\gg (\gamma\mu)^{-1}$, the foreground neutrino loses its identity and blends / equilibrates with the background neutrinos, \ie, $\bm{P}\to\bm{Q}$.

It is interesting to note that the refraction term $\mu \bm{Q}\times\bm{P}$ in Eq.~\eqref{eq:uniform-bg} is due to the foreground neutrino interacting with many background neutrinos in sequence rather than simultaneously as in, \eg, \cite{Friedland:2003dv}. In the weak interaction limit ($\gamma\ll1$), the Bloch vector of the foreground neutrino is rotated only slightly during each interaction with the background neutrinos. These rotations add up coherently because they are indistinguishable. The coherent nature of this term is manifest in the fact that it is proportional to $G_\mathrm{F} n_\nu$. As pointed out in \cite{Friedland:2003eh}, the forward-scattering Hamiltonian in Eq.~\eqref{eq:Vab} captures all the coherent effects of the full Hamiltonian and some of its incoherent effects, although most of the incoherent effects are left out. Here, the incoherent effect in the forward-scattering Hamiltonian stems from the slight entanglement of the foreground neutrino with the background neutrinos that builds up during their interactions. This effect is represented by the ``collision'' term $\bm{C} = \mu\gamma(\bm{Q} - \bm{P})$ in Eq.~\eqref{eq:uniform-bg} that is proportional to $G_\mathrm{F}^2 n_\nu$.%
\footnote{In principle, the slight entanglement among the neutrinos represented by this term can be reproduced as a collision term in the QKE \cite{Blaschke:2016xxt,Froustey:2020mcq}.}
The mean free path resulting from this collision term alone can be estimated to be $\ell_\text{mfp} \sim (\gamma\mu)^{-1} \sim \sigma_\text{wp}^2 / G_\mathrm{F}^2 n_\nu$ if we assume $\delta t\sim \sigma_\text{wp}$ and $\mathcal{V}\sim \sigma_\text{wp}^3$. This agrees with the usual expectation that $\ell_\text{mfp}^{-1} \sim (G_\mathrm{F} E)^2\, n_\nu$ if $\sigma_\text{wp}\sim E^{-1}$, where $E$ is the energy of the neutrino.

\begin{figure}[htb]
    \includegraphics[trim=1 1 1 1, clip]{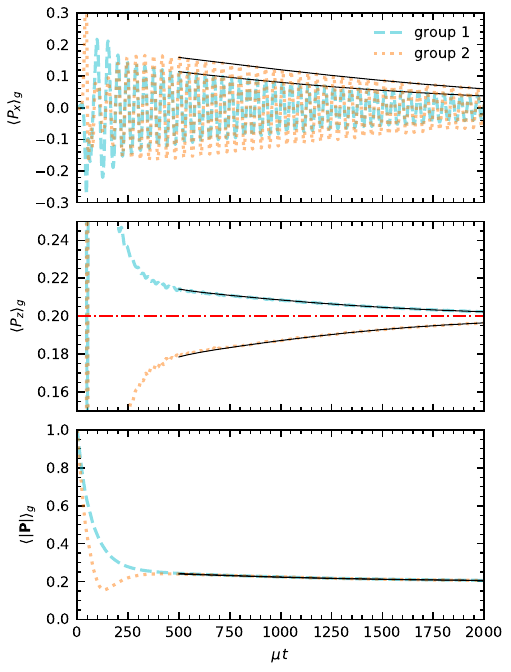}
    \caption{The evolution of an OILE model that consists of 60 $\nu_e$s (group 1) and 40 $\nu_\tau$s (group 2) initially and a small mixing angle $\theta=10^{-3}$. The top two panels show the $x$ and $z$ components of the average Bloch vectors, $\langle \bm{P}\rangle_g$ ($g=1,2$), of the two groups in the mass basis, respectively. The bottom panel shows the average magnitudes of the Bloch vectors. The thin solid curves represent the adiabatic precession / miscidynamic solutions for $|\langle \bm{P}_\perp\rangle_g|$ (top panel), $\langle P_z\rangle_g$ (middle panel), and $\langle |\bm{P}|\rangle_g$ (bottom panel), respectively. The dot dashed line in the middle panel shows the ensemble average $\langle P_z \rangle$ which approximately equals $0.2$ in this case.}
    \label{fig:ex1}
\end{figure}

Also in \cite{Kost:2024esc}, we showed that OILE models with many neutrinos of random velocities can experience sustained collective flavor oscillations over an unexpectedly long period of time when $\gamma\ll1$. (See Fig.~\ref{fig:ex1}.) In this paper, we will demonstrate that this oscillation phenomenon can be described by the adiabatic precession solution in which the Bloch vectors of the neutrinos precess about $\bm{B}$ with a common frequency \cite{Duan:2006an,Duan:2007mv,Raffelt:2007cb,Raffelt:2007xt}. This phenomenon is an example of miscidynamic flavor evolution, in which flavor mixing is in instantaneous equilibrium \cite{Johns:2023jjt,Johns:2024dbe,Johns:2025yxa}. The remainder of the paper is organized as follows. In Sec.~\ref{sec:OILE2} we derive the master equation for the OILE model with stochastic interactions. In Sec.~\ref{sec:prec} we describe the adiabatic precession / miscidynamic solutions and compare them with numerical simulations of the OILE models. Finally, we discuss the implications of our work and conclude in Sec.~\ref{sec:concl}.

\section{OILE model with stochastic interactions}
\label{sec:OILE2}

Consider a ``dilute'' neutrino gas where the probability of the wave packets of more than two neutrinos to overlap simultaneously is negligible. This is a good approximation for astrophysical environments where $n_\nu\lesssim 10^{31}\,\text{cm}^{-3}$ if the size of the neutrino wave packet is $\sigma_\text{wp}\sim 10^{-11}\,\text{cm}$ \cite{Kersten:2015kio}. In \cite{Kost:2024esc}, we assumed that each neutrino interacts with another neutrino during every time step $\delta t$. Here, we assume instead that each neutrino has the same small probability $\mathcal{P}$ to interact during the interval $\delta t$. As in \cite{Kost:2024esc}, we first consider a foreground neutrino with the flavor density operator $\rho_t$ passing through a group of background neutrinos whose flavor quantum states are described by the same density operator $\eta$ before they interact with the foreground neutrino. The density operator of the foreground neutrino after a single interaction with a background neutrino is
\begin{subequations}
\begin{align}
    \rho_{t+\delta t} &= \mathrm{tr}_\eta(e^{-\mathrm{i} V \delta t}\, \rho_t\otimes\eta\, e^{\mathrm{i} V \delta t}) \\
    &= \rho_t -\mathrm{i}\gamma[\eta, \rho_t]
    -\gamma^2(\rho_t - \eta) + \mathcal{O}(\gamma^3),
\end{align} 
\end{subequations}
where $V$ is defined in Eq.~\eqref{eq:Vab} and $\mathrm{tr}_\eta$ denotes the partial trace over the space of the background neutrino. Here we have dropped the geometric factor, which can easily be restored by the replacement $G_\mathrm{F}\to G_\mathrm{F}(1-\vec{v}_a\cdot\vec{v}_b)$.

After interacting with $\Delta N$ different background neutrinos over the interval $\Delta t\ll\omega^{-1}$, where $\omega$ is the vacuum oscillation frequency of the foreground neutrino, the density operator of the foreground neutrino becomes
\begin{widetext}
\begin{align}
    \rho_{t+\Delta t} &= \rho_t + \Delta N \gamma \left\{-\mathrm{i} [{\eta}, {\rho_t}] - \gamma (\rho_t - \eta) + \frac{\gamma}{2} [{\eta}, [{\eta}, {\rho_t}]]\right\} 
    -\frac{(\Delta N\gamma)^2}{2} [{\eta}, [{\eta}, {\rho_t}]] + \mathcal{O}(\gamma^3).
    \label{eq:Delta-rho}
\end{align}
$\Delta N$ for a given $\Delta t$ is a stochastic quantity, but for $\mathcal{P}\ll1$ and $\delta t\ll\Delta t$, $\Delta N$ approximately follows the Poisson distribution, so that
    $\overline{\Delta N^2} = (\overline{\Delta N})^2 + \overline{\Delta N}$,
where the overline denotes the average over the Poisson distribution. Averaging Eq.~\eqref{eq:Delta-rho} over $\Delta N$, we obtain
\begin{align}
    \bar{\rho}_{t+\Delta t} &= \bar{\rho}_t + \mu\Delta t \{-\mathrm{i} [{\eta}, {\bar{\rho}_t}] - \gamma (\bar{\rho}_t - \eta) \}
    -\frac{(\mu\Delta t)^2}{2} [{\eta}, [{\eta}, {\bar{\rho}_t}]] + \mathcal{O}(\gamma^3),
    \label{eq:Delta-rho2}
\end{align}
\end{widetext}
where 
\begin{align}
    \mu = \gamma \left(\frac{\mathcal{P}}{\delta t}\right) = \gamma\, \left(\frac{\overline{\Delta N}}{\Delta t}\right).
\end{align}
Equation~\eqref{eq:Delta-rho2} can be recast as a differential equation for the Bloch vector $\bm{P}(t)=\mathrm{tr}(\bm{\sigma} \bar{\rho}_t)$:
\begin{align}
    \dot{\bm{P}} = (-\omega\bm{B} + \mu\bm{Q})\times\bm{P} + \mu\gamma(\bm{Q} - \bm{P}),
    \label{eq:master-uniform}
\end{align}
where $\bm{Q}=\mathrm{tr}(\bm{\sigma} \eta)$, the vacuum mixing term $-\omega\bm{B}\times\bm{P}$ has been restored, and terms of higher order in $\gamma$ have been dropped. We note that Eq.~\eqref{eq:master-uniform} is the same as that of the original OILE model in \cite{Kost:2024esc} except for the double cross term $-(\gamma\mu/2)\bm{Q}\times(\bm{Q}\times\bm{P})$, which drops out here after averaging over the Poisson distribution. This double cross term does not change the qualitative behavior of the OILE models.

Next, we consider a uniform gas of neutrinos that can be represented by $N\gg1$ (tracing) ``particles'' (which each may represent a collection of real neutrinos with similar flavor evolutions, as in \cite{Kost:2024esc}). The flavor quantum states of these particles can be described by $N$ one-body states, represented by the Bloch vectors $\bm{P}_a$ ($a=1,2,\ldots,N$), because the real neutrinos they represent interact with each other at most once in their lifetimes. In each time step, each particle has the same probability $\mathcal{P}$ to interact with another random particle. Using Eq.~\eqref{eq:master-uniform}, one can show that the flavor evolution of the $a$th particle is governed by the master equation
\begin{align}
    \dot{\bm{P}}_a &= -\omega_a\bm{B}\times \bm{P}_a 
    + \frac{\mu}{N-1}\sum_{b\neq a} \Big[ (1-\vec{v}_a\cdot\vec{v}_b)\bm{P}_b  \times\bm{P}_a
    \nonumber\\
    &\quad
    +\gamma (1-\vec{v}_a\cdot\vec{v}_b)^2(\bm{P}_b - \bm{P}_a)\Big],
    \label{eq:QKE-OILE}
\end{align}
where we have restored the geometric factor $1-\vec{v}_a\cdot\vec{v}_b$. Assuming that the flavor quantum states and the velocities of the particles become uncorrelated with each other after a while, we apply the ansatz \cite{Kost:2024esc}
\begin{align}
    \sum_b (1 - \vec{v}_a\cdot\vec{v}_b)^s \bm{P}_b 
    \approx \langle \bm{P} \rangle \sum_b (1 - \vec{v}_a\cdot\vec{v}_b)^s,
\end{align}
and Eq.~\eqref{eq:QKE-OILE} becomes
\begin{align}
    \dot{\bm{P}}_a &= (-\omega_a \bm{B} + \mu \langle\bm{P}\rangle )\times \bm{P}_a 
    + \frac{4}{3}\mu\gamma(\langle \bm{P}\rangle - \bm{P}_a),
    \label{eq:master}
\end{align}
where $\langle \bm{P} \rangle$ is the ensemble average of the Bloch vectors, and we have also assumed that the velocities of the particles are random and isotropic such that
\begin{align}
    \frac{1}{N} \sum_{b} (1 - \vec{v}_a\cdot\vec{v}_b)^2 \approx \frac{4}{3}.
\end{align}
Again, the double cross term that appears in the original OILE model in \cite{Kost:2024esc}, but does not change the qualitative behavior of the system, no longer appears here in Eq.~\eqref{eq:master}. Averaging Eq.~\eqref{eq:master} over the group of particles $g$ with the same vacuum oscillation frequency $\omega_g$ gives the following equation:
\begin{align}
    \frac{\mathrm{d}}{\mathrm{d}t} \langle \bm{P}\rangle_g &= (-\omega_g \bm{B} + \mu \langle \bm{P}\rangle )\times \langle \bm{P}\rangle_g 
    + \frac{4}{3}\mu\gamma(\langle \bm{P}\rangle - \langle \bm{P}\rangle_g).
    \label{eq:master-g}
\end{align}
Here we use the notation $\langle \cdots \rangle_g$ to denote the average over the particle group $g$ and $\langle \cdots \rangle$ without the subscript for the average over the whole ensemble. 

In Fig.~\ref{fig:ex1}, we show the flavor evolution of an OILE model with 100 particles divided into two groups. The first group has 60 $\nu_e$s initially with the vacuum oscillation frequency $\omega_1=\mu/10$, and the other group has 40 $\nu_\tau$s initially with $\omega_2=\mu/5$. We also adopt a small mixing angle $\theta=10^{-3}$, the impact parameter $\gamma=10^{-2}$, and the probability of interaction per simulation step $\mathcal{P}=0.1$. As shown in the top panel of Fig.~\ref{fig:ex1}, the $x$ components of the average Bloch vectors of both groups $\langle P_x\rangle_g$ ($g=1,2$) oscillate in phase for $\mu t \gtrsim 100$. The $y$ components of the average Bloch vectors $\langle P_y\rangle_{g}$ (not shown) also oscillate in phase with the same frequency. In the bottom panel, one can see that the group average of the magnitudes of the Bloch vectors $\langle |\bm{P}| \rangle_g$ decrease over time due to the collision term. However, the decay rates slow down significantly at $\mu t\gtrsim 280$. In the next section, we will show that this is because the system evolves through a series of quasi-steady states in which the $\langle \bm{P}\rangle_g$ precess about the $z$ axis with a (time-dependent) common frequency.

\section{Adiabatic precession / miscidynamic solution}
\label{sec:prec}

The precession motion $\langle \bm{P}\rangle_g$ in Fig.~\ref{fig:ex1} is similar to that of the mean-field polarization vectors $\bm{F}_{\vec{p}}$ in the neutrino bulb model that leads to spectral swaps / splits \cite{Duan:2006an}. This should not be a surprise since Eq.~\eqref{eq:master-g} is the same as the mean-field equation of motion for the polarization vectors of a homogeneous and isotropic neutrino gas except for the collision term. The flavor evolution of two neutrino groups in the mean-field approximation has been shown to be equivalent to a gyroscopic pendulum in flavor space \cite{Hannestad:2006nj}. For a suitable initial condition, the flavor pendulum can precess about the vertical axis without wobbling \cite{Duan:2007mv}. The precession solution of the mean-field system with a continuous distribution of neutrinos can be solved from a set of self-consistency equations assuming that all polarization vectors rotate about a common axis with the same frequency \cite{Raffelt:2007cb,Raffelt:2007xt}. 

Consider an OILE system of $M$ groups of particles in which there are $N_g \gg 1$ particles with vacuum oscillation frequency $\omega_g$ in the $g$th group ($g=1,\ldots,M$). In the mean-field limit $\gamma \to 0$, the group-averaged Bloch vectors will precess about $\bm{B}$ with the common frequency $\Omega$ if and only if they each are either aligned or antialigned with their Hamiltonian vectors in the frame that rotates about $\bm{B}$ with angular velocity $\Omega$. In this frame, the Hamiltonian vectors are given by
\begin{align}
    \widetilde{\bm{H}}_g = (\Omega -\omega_g)\bm{B} + \mu \bm{D}
\end{align} 
[see Eq.~\eqref{eq:master-g} in the limit $\gamma \to 0$], where
\begin{align}
    \bm{D} = \sum_{g=1}^M f_g \bm{P}^g
\end{align}
with $f_g = N_g / N$ being the fraction of the particles in the group $g$. Here we use $\bm{P}^g$ and $\bm{D}$ for $\langle \bm{P}\rangle_g$ and $\langle\bm{P}\rangle$, respectively, in the precession solutions. The precession ansatz (that Bloch vectors are aligned or antialigned with their Hamiltonian vectors) can be written as
\begin{align}
    \frac{\bm{P}^g}{P^g} = s_g \left(\frac{\widetilde{\bm{H}}_g}{|\widetilde{\bm{H}}_g|}\right),
    \label{eq:ansatz}
\end{align}
where $s_g=\pm1$, and $P^g = |\bm{P}^g|$. 

Following the same procedure outlined in \cite{Raffelt:2007cb}, one obtains the following self-consistency equations in the mass basis:
\begin{subequations}
    \label{eq:prec-con}
\begin{align}
    1 & = \sum_{g=1}^M \frac{f_g s_g P^g}{\sqrt{
        [(\Omega - \omega_g )/\mu + D_z]^2 + D_\perp^2 
    }} , \\
    \Omega &= \sum_{g=1}^M \frac{f_g s_g\omega_g P^g}{\sqrt{
        [(\Omega - \omega_g )/\mu + D_z]^2 + D_\perp^2 
    }}, 
    \label{eq:Omega}
\end{align}
\end{subequations}
where $D_\perp = \sqrt{D_x^2 + D_y^2}$. Given $\omega_g$, $f_g$, $s_g$, $P^g$, $D_z$ and $\mu$, one can solve $\Omega$ and $D_\perp$ from Eq.~\eqref{eq:prec-con}. [$D_z$ is an invariant of Eq.~\eqref{eq:master}.] The components of $\bm{P}^g$ can then be determined using Eq.~\eqref{eq:ansatz}:
\begin{subequations}
    \label{eq:prec-P}
\begin{align}
    P^g_z & = \frac{(\Omega - \omega_g + \mu D_z) s_g P^g}{
        \sqrt{(\Omega - \omega_g + \mu D_z)^2 + (\mu D_\perp)^2}
    }, 
    \label{eq:Pz} 
    \\
    P^g_\perp & = \frac{\mu D_\perp P^g}{
        \sqrt{(\Omega - \omega_g + \mu D_z)^2 + (\mu D_\perp)^2}
    }.
\end{align}
\end{subequations}
This procedure can easily accommodate antineutrinos through the flavor isospin convention \cite{Duan:2005cp}. In this convention, $\bar\nu_e$ ($\bar\nu_\tau$) is represented by a Bloch vector in the $-z$ ($+z$) direction in the flavor basis, and it has the same vacuum oscillation frequency as that of the neutrino of the same energy but with the opposite sign.

The precession solution determined through Eqs.~\eqref{eq:prec-con} and \eqref{eq:prec-P} is a solution to Eq.~\eqref{eq:master-g} for constant $\mu$ in the limit $\gamma\to0$. For a small but finite impact parameter $\gamma$, the OILE model readjusts itself adiabatically to new precession solutions as the collision term slowly reduces $P^g$. From Eq.~\eqref{eq:master-g} one can obtain the following equation that determines $P^g(t)$ in the adiabatic limit:
\begin{align}
    \dot{P}^g = \frac{4}{3}\mu\gamma
    \left(\frac{\bm{D}\cdot\bm{P}^g}{P^g} - P^g\right), \label{eq:Pgdot}
\end{align}
where
\begin{align}
    \bm{D}\cdot\bm{P}^g
    = D_z P^g_z + D_\perp s_g P^g_\perp.
\end{align}

Equations~\eqref{eq:prec-con} through \eqref{eq:Pgdot} are the equation of motion and self-consistency conditions implied by miscidynamics under the adiabatic approximation \cite{Johns:2023jjt,Johns:2024dbe,Johns:2025yxa}. Miscidynamics is a coarse-graining of neutrino quantum kinetics that follows from the approximation that the coarse-grained---in this case group-averaged---system is in instantaneous mixing equilibrium: $\widetilde{\bm{H}}_g \times \bm{P}^g \approx 0$ at all times. The equilibrium condition is satisfied in a time-dependent rotating frame \cite{Johns:2025yxa}. We use the terms ``precession solution'' and ``miscidynamic solution'' interchangeably because they coincide for this system.

We calculated the adiabatic precession solution for the two-group OILE model in Sec.~\ref{sec:OILE2} with initial values of $P^g$ and $s_g$ taken from the numerical simulation at $\mu t = 500$. The results are shown as thin solid curves in Fig.~\ref{fig:ex1}, and they agree well with the numerical simulations. In the bottom panel of Fig.~\ref{fig:ex1}, one can see that the decay of the group-averaged magnitudes of the Bloch vectors $\langle |\bm{P}|\rangle_g$ slows significantly as the system enters the quasi-steady precession state at $\mu t\sim 280$. This can be understood as follows. In the early stage of evolution ($\mu t\lesssim \gamma^{-1}=100$), the neutrino gas develops kinetic decoherence in the flavor space as in the mean-field limit ($\gamma=0$), \ie, the Bloch vectors of the neutrinos with the same energy but different velocities diverge from each other \cite{Raffelt:2007yz}. This leads to quantum decoherence on the timescale of $\sim(\gamma\mu)^{-1}=100/\mu$ due to the collision term $\frac{4}{3}\gamma\mu (\langle \bm{P}\rangle - \bm{P}_a)$ in Eq.~\eqref{eq:master}. After the system enters the quasi-steady precession state, the Bloch vectors of the neutrinos become very similar to each other, and the rate of quantum decoherence slows down significantly. 

\begin{figure*}[htb]
    \includegraphics[trim=1 1 1 1, clip]{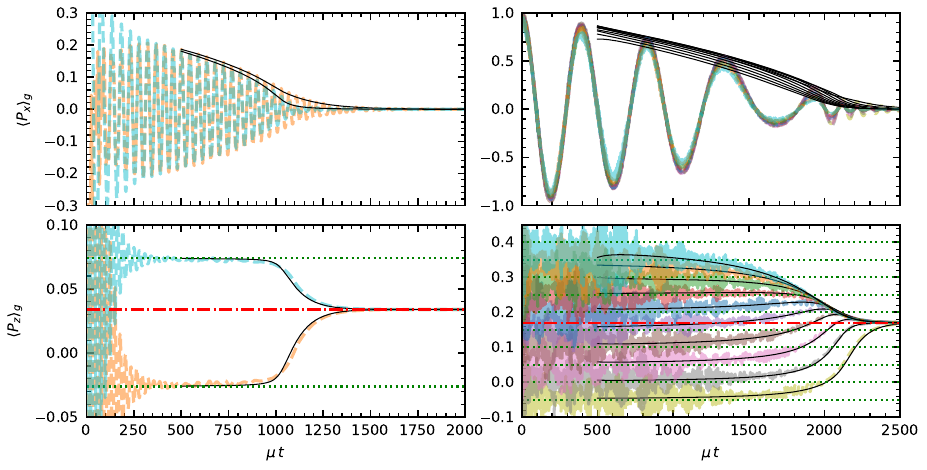}
    \caption{Similar to Fig.~\ref{fig:ex1} but for two different OILE models with a large mixing angle $\theta=0.7$. The horizontal dotted lines in the lower panels are the approximate values of $P^g_z$  in the precession / miscidynamic solution according to Eq.~\eqref{eq:stage2-approx}. See the text for details.}
    \label{fig:ex23}
\end{figure*}

In Fig.~\ref{fig:ex23}, we show the flavor evolution of two additional OILE models. The model shown in the left panels is the same as the one shown in Fig.~\ref{fig:ex1} except it has a much larger mixing angle $\theta=0.7$. An interesting feature of this model is that the $\langle P_z\rangle_g$ remain approximately constant after the system enters the quasi-steady precession state until $\mu t \gtrsim 1000$. This is because the Bloch vectors are nearly in the transverse directions initially. Then, the magnitudes of the Bloch vectors decrease on the timescale $(\gamma\mu)^{-1}=100/\mu$ until they become similar. As a result,
\begin{align}
    P^g \approx D_\perp\gg |D_z|,\, |\omega_g/\mu|
    \label{eq:Pg-approx}
\end{align}
when the system first enters the quasi-steady precession state. According to Eqs.~\eqref{eq:Omega} and \eqref{eq:Pz}, 
\begin{subequations}
\label{eq:stage2-approx}
    \begin{align}
    \Omega &\approx \sum_g f_g s_g \omega_g
    \intertext{and}
    P^g_z &\approx \left(\frac{\Omega - \omega_g}{\mu} + D_z \right) s_g
\end{align}
\end{subequations}
are approximately constant in this stage. This approximation breaks down when $P^g_\perp$ becomes comparable to $|P^g_z|$. Afterwards, $\bm{P}^g$ quickly approaches $D_z \bm{B}$, which is the true equilibrium state of the system. We plot the approximations of $P^g_z$ according to Eq.~\eqref{eq:stage2-approx} in the lower left panel of Fig.~\ref{fig:ex23}, and they agree with the numerical results until $\mu t\sim 1000$.

The model shown in the right panels of Fig.~\ref{fig:ex23} also has the large mixing angle $\theta=0.7$, but with 10 groups of particles, all with Bloch vectors $(0,0,1)$ initially. The $g$th group ($g=1,\ldots,10$) has $N_g = 20$ (30) particles if $g$ is odd (even) with the vacuum oscillation frequency $\omega_g = (-0.3 + 0.05 g)\mu$. The behavior of this model is similar to the previous one, although Eq.~\eqref{eq:stage2-approx} is not a good approximation for the first few particle groups because their $P^g_z$ are not much smaller than $P^g_\perp$ when the system enters the quasi-steady precession state.

\section{Discussion and conclusion}
\label{sec:concl}
We have derived the master equations for the OILE model with stochastic interactions. As in the original OILE model in \cite{Kost:2024esc}, the neutrino gas described by the new model evolves as the mean field for $ t\lesssim (\gamma\mu)^{-1}$, where $\mu$ and $\gamma$ characterize the strengths of forward scattering of the neutrinos at the mean-field and wave packet levels, respectively. Flavor decoherence through quantum entanglement among neutrinos, which is represented by the collision terms in the master equations, becomes important for $t\gtrsim (\gamma\mu)^{-1}$. In the limit $\gamma\ll1$, which is expected in astrophysical environments such as CCSNe and NSMs \cite{Kost:2024esc}, the combination of the kinematic flavor decoherence on the mean-field level and the quantum flavor decoherence on the wave packet level drives the neutrino gas into a quasi-steady state similar to the precession mode of collective neutrino oscillations in the mean-field limit \cite{Duan:2006an,Duan:2007mv,Raffelt:2007cb,Raffelt:2007xt}. In this quasi-steady state, the flavor Bloch vectors of the neutrinos rotate about a common axis with a common frequency in flavor space. Afterwards, the collision term continues to depolarize the flavor Bloch vectors of the neutrinos and eventually drives them to the true stationary state, in which flavor equilibration is complete.

Three differences from the previous work on the mean-field models are worth emphasizing. First, quasi-steady precession states emerge in the OILE models at the coarse-grained (group-averaged) level despite the presence of fine-grained fluctuations. Second, the OILE models exhibit equilibration to a quasi-steady state, not only instantaneous-equilibrium tracking. Third, the adiabatic approximation is successful despite the nonconservation of $P^g$. The adiabatic evolution through different precession states is in fact driven organically by the collision term rather than ``externally'' by a slowly changing parameter in the Hamiltonian (as in \cite{Raffelt:2007cb}, where the neutrino density is decreased by hand). These aspects of the flavor evolution were all anticipated in \cite{Johns:2023jjt} and further elaborated on in \cite{Johns:2024dbe,Johns:2025yxa}.

The flavor evolution of the OILE models described above is an example of miscidynamic flavor evolution \cite{Johns:2023jjt,Johns:2024dbe,Johns:2025yxa}. The key idea of miscidynamics is that the flavor evolution of a neutrino gas can be solved on the macroscale $\ell_\text{mfp}$ if the neutrino oscillations on the much smaller mesoscale $\ell_\text{osc}$ can be described as a series of ``mixing equilibria''. Indeed, we were led to the precession solution presented in this work by applying adiabatic miscidynamics to Eq.~\eqref{eq:master-g}. The mixing equilibria in this case are the quasi-steady precession states characterized by neutrino oscillations on the mesoscale $\ell_\text{osc}\sim \omega^{-1}$, which is much smaller than the macroscale $\ell_\text{mfp}\sim (\mu\gamma)^{-1}$ in CCSNe and NSMs where the impact parameter can be as small as $\gamma\sim 10^{-10}$ \cite{Kost:2024esc}. The miscidynamic approach can also be applied to other systems, if the mixing equilibria can be determined beforehand.

\begin{acknowledgments}
We thank G.~Raffelt for helpful discussions.
A.~K.\ and H.~D.\ are supported by the US DOE NP grant No.\ DE-SC0017803 at UNM. L.~J.\ is supported by the US Department of Energy and Los Alamos National Laboratory under contract 89233218CNA000001 and by a Feynman Fellowship through LANL LDRD project No.\ 20230788PRD1.
\end{acknowledgments}

\bibliographystyle{apsrev4-2}
\bibliography{oile2}

\end{document}